\documentclass{phb-proc4-auth}

\usepackage{graphicx}
\usepackage{amssymb}

\begin{document}
\begin{frontmatter}


\journal{SCES '04}


\title{ Mechanism for lifting the degeneracy \\
in the double-exchange spin ice model on a kagom\'e lattice:\\ 
Dodecamer formation}

%
%
%
%
%
%

\author{Yoshihiro Shimomura\thanksref{shimo}\corauthref{1}}, 
\author{Shin Miyahara} and 
\author{Nobuo Furukawa}

%
 
\address{Department of Physics and Mathematics, Aoyama Gakuin University, 
Sagamihara, Kanagawa 229-8558, Japan}

%
%
%
%

\thanks[shimo]{This work was supported by a Grant-in-Aid for 21st COE Program
from the Ministry of Education, Culture, Sports, Science 
and Technology of Japan.}

%
%
%
%

\corauth[1]{
Corresponding Author: Department of Physics and Mathematics, 
Aoyama Gakuin University, 
5-10-1 Fuchinobe, Sagamihara, 229-8558, Japan. 
Phone: +81-42-759-6288, 
Fax: +81-42-759-6542, 
Email: shimomura@phys.aoyama.ac.jp 
(Y. Shimomura)}


\begin{abstract}


We investigated the double-exchange spin ice model on a kagom\'e 
lattice by Monte Carlo simulation in order to study a mechanism 
for lifting the degeneracy in frustrated electron systems. 
We show specific heat and vector spin chirality data 
on a finite lattice. 
Specific heat has a double-peak structure: 
A broad peak and a sharp peak are at 
$k_{\rm B}T/t \sim 0.15$ and 0.015, respectively, 
where $t$ is the transfer integral of electrons. 
The broad peak corresponds to 
a crossover to the spin ice-like state, 
on the other hand, the sharp one a transition 
to a dodecagonal spin cluster (dodecamer) state. 
We discuss the interplay 
between the formation of the dodecamer state 
and the lifting of the macroscopic degeneracy. 

\end{abstract}

%
%



\end{frontmatter}

%
%
%
%
%
%

Macroscopic degeneracy is one of the well-known properties 
in frustrated systems. 
However, such a degeneracy should be lifted at $T=0$ 
in realistic systems because of the third law of thermodynamics. 
Considering these circumstances, 
it is significant to investigate mechanisms to lift the degeneracy. 
In particular, a peculiar mechanism is expected 
in electron systems with the frustration: 
The motion of electrons plays an important role 
for lifting macroscopic degeneracy, 
which may lead to some kinds of exotic phases. 

In order to see such a mechanism, 
we investigate a double-exchange spin ice (DESI) model 
on a kagom\'e lattice 
by a Monte-Carlo (MC) simulation \cite{Shimomura_04}, 
where the kagom\'e lattice consists of corner-shared 
up- and down-triangles. 
The DESI model is the Anderson-Hasegawa model \cite{Anderson-Hasegawa_55} 
with a following condition: 
Localized spins have uniaxial anisotropies that 
they are forced to point either inward (in-spin) or 
outward (out-spin) for an up-triangle. 
In this case, localized spins are coplanar. 
The uniaxial anisotropies and 
ferromagnetic interaction between the nearest-neighbor (n.n.) spins 
due to the Anderson-Hasegawa mechanism 
cause the frustration in the model 
for the same reason seen in the spin ice systems \cite{Bramwell_01}. 

\begin{figure}[h]
\begin{center}
\includegraphics[width=5cm,keepaspectratio]{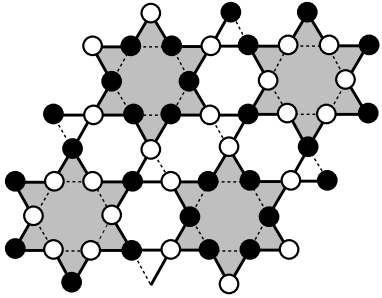}
\end{center}
\caption{
Dodecamer formation in the kagom\'e lattice: 
The shadowed dodecagonal spin cluster is a dodecamer, 
where in-(out-)spins for up-triangles 
are represented by open (closed) circles. 
There are two types of bond in the DESI system. 
One is a bond which connects an in-spin and an out-spin (solid line), 
and the other is that which connects 
the n.n. pairs of the same spins (dashed line). 
The electron transfer integral has larger value 
on the former bond than the latter. 
The dodecamer consists of twelve spins connected by solid bonds 
along which electrons can move easily. 
The dodecamer order is a bond order rather than a spin order.
The dodecamer state means translational symmetry breaking. 
}
\label{fig:ddcmr}
\end{figure}

From our previous MC calculations, 
following results have been obtained \cite{Shimomura_04}: 
At sufficiently low temperatures, 
a dodecagonal localized spin cluster, ``dodecamer'' 
[see Fig. \ref{fig:ddcmr}], 
is realized in a wide doping region 
$n \simeq 1/3 \sim 1/2$, where $n$ is the number of particles per site.
The dodecamer order is driven by 
both the kinetic energy gain and the frustration. 

In this paper, specific heat and vector spin chirality data of the system 
are shown, in wide temperature range 
in order to investigate a relation 
between the entropy release and the spin structure. 
According to the results obtained here, it is found that 
the system undergoes spin ice-like states 
as in the spin ice systems \cite{Bramwell_01} 
at intermediate temperatures 
and the dodecamer state at further low temperatures. 

We have performed MC calculations \cite{Yunoki_98} 
to study the thermodynamics of 
the system at finite temperatures. 
We typically run 100,000 MC steps 
for measurement after 10,000 thermalization steps. 

\begin{figure}[h]
\begin{center}
\includegraphics[width=8cm,keepaspectratio]{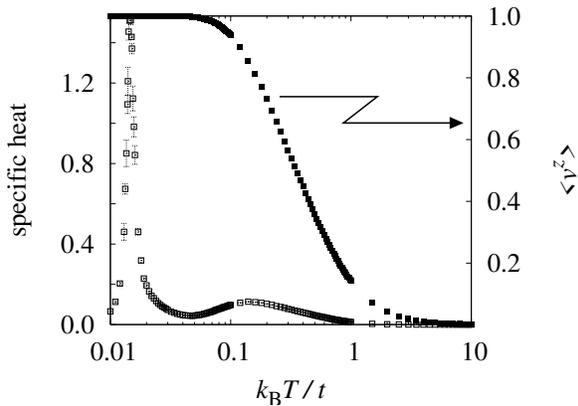}
\end{center}
\caption{
Temperature dependence of the specific heat and 
a vector spin chirality $\langle v^{z} \rangle$ 
for 6 x 6 unit cells (= 108 sites) and the chemical potential $\mu = 0$. 
Error bars are within the sizes of the symbols. 
Specific heat shows a double-peak structure: 
One is at $k_{\rm B}T/t \sim 0.015$ and 
the other $k_{\rm B}T/t \sim 0.15$. 
$\langle v^{z} \rangle$ changes rapidly 
at around the broad peak of the specific heat data. 
Furthermore, the sharp peak exists 
in the $\langle v^z \rangle \simeq 1$ region. 
}
\label{fig:spe_chi}
\end{figure}

In order to see how the entropy is released in the system, 
specific heat was investigated, 
and data are shown in Fig. \ref{fig:spe_chi} 
for $6 \times 6$ unit cells (= 108 sites) 
with the chemical potential $\mu$ = 0. 
We can see a double-peak structure: A broad peak is seen 
at $k_{\rm B}T/t \sim 0.15$ and 
a sharp peak is at $k_{\rm B}T/t \sim 0.015$, 
where $t$ is the transfer integral of electrons. 

To clarify the origin of the broad peak, 
we check a spin configuration of the system. 
A vector spin chirality is defined for triangles in the kagom\'e lattice 
which has the form, 
$
\vec{v} = -\frac{2}{\sqrt{3}}\left(
  \vec{S}_{i}\times \vec{S}_{j} 
+ \vec{S}_{j}\times \vec{S}_{k}
+ \vec{S}_{k}\times \vec{S}_{i} \right), 
$
where $i$, $j$ and $k$ represents three sites of a triangle. 
Here, we set $|\vec{S}_i| \equiv 1$. 
We only consider the $z$-component of $\vec{v}$, 
because localized spins are coplanar in the system.
Each triangle is allowed to take 
``three-out'' ($v^{z} = -3$), ``three-in'' ($v^{z} = -3$), 
``two-in one-out'' ($v^{z} = 1$) 
or ``one-in two-out'' ($v^{z} = 1$) spin structures. 
At sufficiently high temperatures, 
all of these patterns are realized equally, i.e., 
$\langle v^{z}\rangle = 0$, 
since both one-in two-out and two-in one-out have 
a three-fold degeneracy. 
On the other hand, as the temperature is lowered, 
three-out and three-in states are excluded, 
because they have higher energy than the others 
due to the n.n. ferromagnetic interaction. 
Then, the system has $\langle v^{z} \rangle \rightarrow 1$. 
This situation is similar to the spin ice systems \cite{Bramwell_01}. 
Thus, we call such states an ``ice state'', hereafter.

Temperature dependence of $\langle v^{z} \rangle$ 
are shown in Fig. \ref{fig:spe_chi}. 
$\langle v^{z} \rangle$ grows as the temperature is lowered, 
and almost saturates below the broad peak. 
This means that the system becomes the ice state 
at the broad peak temperature. 
Thus, we conclude that the broad peak of the specific heat data 
corresponds to a crossover to the ice state. 
Note that the ice state still has the macroscopic degeneracy 
due to the frustration. 

Next, let us consider the sharp peak in the specific heat data. 
We investigated a dodecamer structure factor $D_{q}$ 
in Ref \cite{Shimomura_04}. 
The largest peak of $D_q$ is defined as $\bar{D}_q$, 
which shows a rapid change at $k_{\rm B}T/t \sim 0.015$. 
Temperature dependence of $\bar{D}_q$ indicates 
that there is an inflection point in this temperature range.  
The sharp peak temperature is almost consistent with 
that of the inflection point. 
Thus, it is considered that 
the sharp peak corresponds to a phase transition 
from the ice state to the dodecamer state, 
which is smeared out by the finite-size effect.  

In conclusion, the process for lifting the degeneracy 
in the system is as follows: 
At intermediate temperatures, 
the n.n. ferromagnetic interaction leads the system to the ice state,
which still has the macroscopic degeneracy.
According to this result, the short-range correlation 
is considered to be insufficient to lift the degeneracy, 
often seen in the classical spin systems. 
Furthermore, at sufficiently low temperature region, 
the effective long-range interaction is relevant 
in order to gain the kinetic energy, 
and the degeneracy is lifted by the dodecamer formation 
accompanied with the translational symmetry breaking. 

%
%
%
%

%
%
%
%


\end{document}